\documentclass[twocolumn,showpacs,preprintnumbers,amsmath,amssymb]{revtex4-1} \usepackage{dcolumn}

\usepackage{graphicx} \usepackage{bm}

\begin{document}
\title{Damping of electron Zitterbewegung in carbon nanotubes}
\date{\today}
\author{Tomasz M. Rusin$^1$}

\email{Tomasz.Rusin@orange.com}

\author{Wlodek Zawadzki$^2$}
\affiliation{$^1$Orange Customer Service sp. z o. o., Al. Jerozolimskie, 02-326 Warsaw, Poland\\
             $^2$ Institute of Physics, Polish Academy of Sciences, Al. Lotnik\'ow 32/46, 02-688 Warsaw, Poland}

\pacs{72.80.Vp, 42.50.-p, 41.75.Jv, 52.38.-r}

\begin{abstract}
Zitterbewegung (ZB, trembling motion) of electrons in semiconductor carbon nanotubes is described taking into account
dephasing processes. The density matrix formalism is used for the theory. Differences between
decay of ZB oscillations due to electron localization and that due to dephasing are discussed.
\end{abstract}
\maketitle

\section{Introduction}

The phenomenon of Zitterbewegung (ZB, trembling motion) was devised by Schrodinger~\cite{Schroedinger1930},
who observed that, if one uses the Dirac equation for free relativistic electrons in a vacuum, the velocity operator
does not commute with the Dirac Hamiltonian. This means that the resulting electron velocity is not a
constant of the motion even in the absence of external fields. It was later realized that the appearance of ZB in the
Dirac equation is a result of the two-band structure of its energy spectrum~\cite{BjorkenBook}. Lock~\cite{Lock1979}
observed, that since the ZB had been predicted for plane waves, it was not clear what the trembling meant for
electron uniform distribution in space. Lock pointed out that, if the electron is represented in space by a localized
Gaussian wave packet, its ZB oscillations decay in time as a consequence of the Riemann-Lebesgue lemma.
In 2010 Gerritsma {\it et al.}~\cite{Gerritsma2010} succeeded in simulating the 1+1 Dirac equation with the resulting
Zitterbewegung using cold ions interacting with laser beams. The ZB of charge carries was also predicted in superconductors
and narrow-gap semiconductors as a consequence of two-band energy spectra in such materials. Since 2005, when papers
by Zawadzki~\cite{Zawadzki2005} and Schliemann {\it et al.}~\cite{Schliemann2005} appeared, the trembling motion in
narrow-gap materials and similar periodic systems became an intensively studied subject, as reviewed in Ref.~\cite{Zawadzki2011}.

Our purpose in the present work is to introduce to the description of ZB dephasing processes. This requires
the density matrix (DM) formalism which, to our knowledge, has not been attempted in the literature for the ZB problem. We
choose for our description single-wall semiconductor carbon nanotubes (CNT) since they are characterized by a
relatively simple~$2\times 2$ electron Hamiltonian for which majority of calculations can be carried out analytically.
The phenomenon of ZB in CNT was treated in the past with the use of~Heisenberg time-dependent
operators~\cite{Zawadzki2006,Rusin2007b} or by solving the Schrodinger equation
with the time-dependent Hamiltonian~\cite{Rusin2013b}. This gives us a possibility
to check that, as the damping processes are assumed to vanish, the different formalisms give similar results.

\section{Theory and results}

Within the~${\bm k}\cdot {\bm p}$ theory in the absence of external fields the electron
Hamiltonian at the~$K$ point of the Brillouin zone of CNT is~\cite{Ando1993}
\begin{equation} \label{HCNT}
\hat{H} = u\hbar \left(\begin{array}{cc} 0 & -i\kappa_{n\nu} - k \\ i\kappa_{n\nu} -k & 0 \end{array} \right),
\end{equation}
where~$u\simeq 10^6$ cm/s,~$k$ is the wave vector in the~$y$ direction parallel to the tube's
axis,~$\kappa_{n\nu}=(2\pi/L)(n-\nu/3)$,~$n=0, \pm 1,...$, and~$L$ is the
circumference length of CNT. For metallic CNT~$\nu=0$, while for semiconductor CNT~$\nu=\pm 1$.
The eigenenergies of~$\hat{H}$ are~$E_{1,2}=\pm u\hbar\xi$, where~$\xi=\sqrt{k^2+\kappa_{n\nu}^2}$,
and the eigenfunctions are:~$w_1=[(i\kappa_{n\nu}+k)/\xi,-1]/\sqrt{2}$
and~$w_2=[(i\kappa_{n\nu}+k)/\xi,+1]/\sqrt{2}$, respectively.
The two signs in the energy and the wave functions correspond to the conduction and valence subbands, respectively.
Since the Hamiltonian
is a~$2\times 2$ operator, the density matrix~$\hat{\rho}$ of the system can be obtained in
the well-known two-level formalism~\cite{BoydBook}. In the presence of damping, the
Liouville equation for~$\hat{\rho}$ is
\begin{equation} \label{Liou}
 \frac{d\hat{\rho}}{dt} = \frac{-i}{\hbar}(\hat{H}\hat{\rho} - \hat{\rho}\hat{H}) - \hat{D}(\hat{\rho}),
\end{equation}
in which~$\hat{D}(\hat{\rho})$ describes phenomenologically dephasing processes taking place
during the electron motion. Matrix elements of~$\hat{D}(\hat{\rho})$ between eigenvectors~$w_1$ and~$w_2$
are taken in the standard form~\cite{BoydBook}
\begin{eqnarray} \label{LiouDump}
 \hat{D}(\hat{\rho})_{mn} = \gamma_{mn} \langle w_m|\hat{\rho}-\hat{\rho}^{eq}|w_n\rangle \hspace*{1em} (m,n=1,2),
\end{eqnarray}
where~$\hat{\rho}^{eq}$ is DM at equilibrium, when the electron is in the
state with negative energy~$E_2$. Then~$\rho^{eq}_{22} = 1$
and other~$\rho^{eq}_{mn}$ are zero. The constants~$\gamma_{11}=\gamma_{22}=1/T_1$
describe the relaxation of electron population excited to the conduction subband,
while~$\gamma_{12}=\gamma_{21}=1/T_2$ describe the decay of coherence between the upper and lower electron states.
We disregard long-time relaxation processes with~$T > 2000$~fs, as observed in CNTs~\cite{Kono2003}.

By calculating the matrix elements of both sides in Eq.~(\ref{Liou}) one obtains
four first-order differential equations for the matrix elements of~$\hat{\rho}$.
We have for instance:~$\dot{\rho}_{12}=-(2i\omega+1/T_2){\rho}_{12}$, where~$\omega=u\xi$.
Solving these equations we obtain
\begin{equation} \label{rho}
 \hat{\rho}(t) = \left(\begin{array}{ll}
     c_{11} e^{-t/T_1}           & c_{12} e^{-2i\omega t-t/T_2} \\
     c_{21} e^{2i\omega t-t/T_2} & 1-c_{11} e^{-t/T_1} \end{array} \right),
\end{equation}
where the coefficients~$c_{mn}$ must be determined from initial conditions. We assume that at~$t=0$
the electron is represented by a Gaussian wave packet having one non-zero
component:~$F_{n\nu}(k) = (1,0)^T g_{n\nu}(k)$, with
\begin{equation} \label{Gauss}
 g_{n\nu}(k) = 2\sqrt{\pi d} e^{-d^2k^2/2}.
\end{equation}
The packet is normalized to~$\int_{-\infty}^{\infty}|g_{n\nu}(k)|^2 dk=2\pi$.
We assume that it consists of states belonging to one pair of energy subbands with
given~$n$ and~$\nu$, see Eq.~(\ref{HCNT}). In the following we drop the subscript~$n\nu$.
The packet is a combination of states having positive and negative
energies:~$F(k)=a_1w_1g(k) + a_2w_2 g(k)$ with~$a_1=a_2=(k-i\kappa_{n\nu})/\sqrt{2}\xi$.
Thus the probability densities are~$p_1=|a_1|^2=|g(k)|^2/2$
and~$p_2=|a_2|^2=|g(k)|^2/2$, so that the states with positive and negative
energies contribute the same probabilities to the initial packet.
The density matrix corresponding to~$F(k)$ is
\begin{equation} \label{rhoF}
 \hat{\rho}_F = F(k) \cdot F(k)^{\dagger} = |g(k)|^2 \left(\begin{array}{ll} 1 & 0 \\ 0 & 0\end{array} \right).
\end{equation}
The matrix~$\hat{\rho}_F$ is given in the representation of the upper and lower components.
To find~$\hat{\rho}_F$ in the representation of the eigenstates of~$\hat{H}$ we calculate
the matrix elements of~$\hat{\rho}_F$ between the states~$w_1$ and~$w_2$. For DM at~$t=0$ one obtains
\begin{figure}
\includegraphics[width=8.0cm,height=8.0cm]{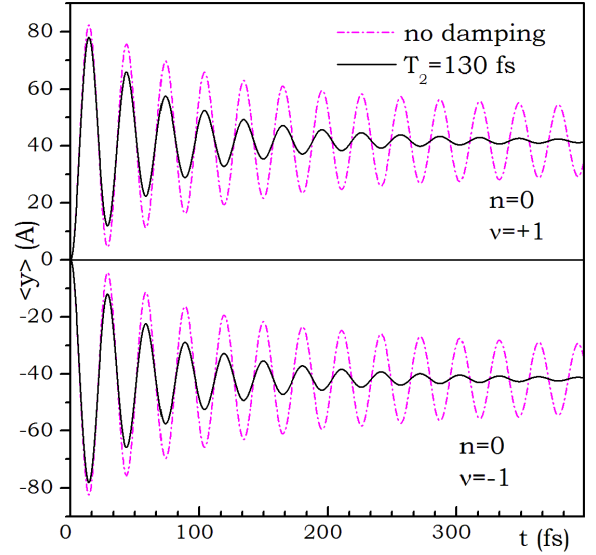}
\caption{Calculated average position of electron Gaussian wave packet for two pairs of subbands in a carbon nanotube versus time.
         Solid lines: damping included. Dash-dotted lines: no damping. Packet width is~$d=160 \AA$, tube circumference is~$L=200 \AA$,
         initial packet momentum is zero.} \label{FigRhoCNT}
\end{figure}
\begin{equation} \label{rho0}
 \hat{\rho}(0) =\frac{1}{2}|g(k)|^2 \left(\begin{array}{cc} 1 &1 \\ 1 & 1 \end{array} \right),
\end{equation}
which gives:~$c_{11}=c_{12}=c_{21}=|g(k)|^2/2$, see Eq.~(\ref{rho}).
The velocity operator in the representation of upper and lower components
is~$\hat{v} = (\partial {\hat H})/(\partial \hbar k) = -u\sigma_x$.
Its counterpart in the representation of the eigenstates of~$\hat{H}$ is
\begin{equation} \label{V}
 \hat{V} = \frac{u}{\xi} \left(\begin{array}{cc} k &i\kappa_{n\nu} \\ -i \kappa_{n\nu}& -k \end{array} \right).
\end{equation}
The average velocity is, see remark~\cite{Note2}
\begin{eqnarray} \label{v11}
 \langle v(t)\rangle &=& {\rm Tr}(\hat{V}\hat{\rho}) = -u\int_{-\infty}^{\infty}
              \frac{\kappa_{n\nu}}{\xi}\sin(2\omega t)e^{-t/T_2}|g(k)|^2 dk \nonumber \\
                     &+& u\int_{-\infty}^{\infty} \frac{k}{\xi}(e^{-t/T_1}-1)|g(k)|^2 dk,
\end{eqnarray}
and the average position is
\begin{equation} \label{y11}
\langle y(t)\rangle = \int_0^t\langle v(t')\rangle dt',
\end{equation}
where the initial condition is:~$y(0)=0$. Equations~(\ref{v11}) and~(\ref{y11}) describe the motion of
electron packet in the presence of dephasing processes which modify the motion in two different ways.
The decay of coherence between electron states in the conduction and valence subbands,
as characterized by~$T_2$, alters the ZB oscillations, while the relaxation of electrons
from the upper to lower energy subbands, as characterized by~$T_1$, changes the
rectilinear part of the motion. To analyze quantitatively the effect of damping
we calculate the average packet position for the Gaussian packet~(\ref{Gauss})
assuming that the times~$T_1$ and~$T_2$ do not depend on~$k$,~$n$ and~$\nu$.
We take~$T_2=130$~fs, as determined experimentally in Ref.~\cite{Voisin2003}.
For the packet of Eq.~(\ref{Gauss}) the last integral in Eq.~(\ref{v11}) vanishes and
the relaxation time~$T_1$ is of no relevance to our problem.

In Fig.~1 we show the calculated results taking the
packet width~$d=160 \AA$ and the tube circumference~$L=200 \AA$.
These parameters were used in Ref.~\cite{Rusin2007b}.
We consider a semiconductor CNT and the wave packet of states with~$n=0$ and~$\nu=\pm 1$ subbands.
In the absence of damping the oscillations decay as~$t^{-1/2}$ (dash-dotted lines),
while in its presence they decay exponentially with the characteristic time~$T_2$ (solid lines).
In the presence of damping the decay of oscillations can be significantly
faster which would make experimental observations of ZB more difficult.

To analyze the origin of the oscillating and rectilinear terms in Eq.~(\ref{v11}) we decompose
DM in Eq.~(\ref{rho}) into two matrices:~$\hat{\rho} = \hat{\rho}_r + \hat{\rho}_Z$,
where~$\hat{\rho}_r=\left(\begin{array}{cc} \rho_{11} &0 \\ 0 & \rho_{22} \end{array} \right)$ and
      $\hat{\rho}_Z=\left(\begin{array}{cc} 0 & \rho_{12} \\ \rho_{21} & 0 \end{array} \right)$.
Then the average velocity is also a sum of two terms
\begin{eqnarray}
\label{vr}
\langle v_r(t)\rangle &=& {\rm Tr}(\rho_r V)= u\int_{\-\infty}^{\infty} \frac{k}{\xi}(e^{-t/T_1}-1)|g(k)|^2 dk, \ \ \ \\
\label{vZ}
\langle v_Z(t)\rangle &=&{\rm Tr}(\rho_Z V) = \nonumber \\
     &=&-u\int_{-\infty}^{\infty} \frac{\kappa_{n\nu}}{\xi}\sin(2\omega t)e^{-t/T_2}|g(k)|^2 dk.
\end{eqnarray}
The above equations relate ZB oscillations to the off-diagonal elements of DM and rectilinear motion
to the diagonal elements.

As mentioned in the Introduction, a similar problem of ZB oscillations in CNT in the absence of damping
was considered in Ref.~\cite{Rusin2007b} with the use of Heisenberg picture. In this approach,
one calculates:~$\langle v(t)\rangle =u \langle F| e^{i\hat{H}t/\hbar} \sigma_x e^{-i\hat{H}t/\hbar}|F\rangle$.
To compare our results with those of Ref.~\cite{Rusin2007b}
we set in Eq.~(\ref{v11})~$T_1,T_2\rightarrow \infty$, which gives
\begin{equation} \label{v11_0}
 \langle v(t)\rangle = -u\int_{-\infty}^{\infty} \frac{k_{n\nu}}{\xi}\sin(2\omega t)|g(k)|^2 dk.
\end{equation}
This agrees, up to the sign, with~$\langle v(t)\rangle$ obtained in Ref.~\cite{Rusin2007b}, in which
the velocity operator at~$t=0$ had an opposite sign.

\section{Discussion and summary}

First, we want to comment on the results presented in Fig.~1. At the first sight, the decays
of ZB oscillations without and
with the damping differ only quantitatively. However, one should bear in mind that the decay without the damping is
technically caused by the fact that we used a Gaussian wave packet to localize the electron in space. In this case the
amplitude of ZB diminishes as~$t^{-1/2}$ and this decay is caused by different velocities of interfering sub-packets
of positive and negative energies, which progressively cease to overlap, see Ref.~\cite{Rusin2007b}. The choice of localizing
packet is somewhat arbitrary and, had we chosen a different packet, the decay would have also been different.
In particular, when there is no localization by the packet, the ZB oscillations do not diminish in time,
see Ref.~\cite{Zawadzki2006}.
On the other hand, the decay due to damping is a real physical effect caused by decoherence and relaxation processes.
The dephasing processes can be of various kinds: the electron can be scattered elastically or non-elastically,
or it can loose its energy by emitting radiation. We do not describe these processes, they are summarized
phenomenologically by the times~$T_1$ and~$T_2$. Still, we want to mention that, since the electron wave packet
includes the states of both positive and negative electron energies, i.e. the electron is not in a single eigenstate of
the Hamiltonian~(\ref{HCNT}), it can emit radiation.

In Eq.~(\ref{Gauss}) we assumed a symmetric wave packet centered around the value~$k_0=0$. If we assumed~$k_0\neq 0$,
the second integral in Eq.~(\ref{v11}) would not vanish, which would give a rectilinear (classical) component
to the motion. However, as discussed in Ref.~\cite{Rusin2013b}, it is not clear how to furnish to the electron
a sizable non-vanishing momentum~$\hbar k_0$, so we do not discuss this case here.

We assumed above that at~$t=0$ the wave packet consists of states belonging to
one pair of energy subbands, but in general it can be a combination of states belonging to more pairs of subbands.
In this case~$g(k)$ in Eq.~(\ref{Gauss}) should be replaced by~$\sum_{n\nu}g_{n\nu}(k)$
and in Eqs.~(\ref{v11}) --~(\ref{vZ}) one should perform additional summations over~$n$ and~$\nu$.
However, as shown in Ref.~\cite{Rusin2013b},
shapes and parameters of realistic wave packets depend on physical mechanisms used for their creation.
Thus the determination of~$g_{n\nu}(k)$ sub-packets is out of scope of the present work.
On the other hand, as pointed out in Ref.~\cite{Rusin2013b}, a single ultra-short laser pulse can create
an oscillating wave packet consisting mainly of states belonging to one pair of energy subbands, which justifies the
approach used above.

Finally, it can be seen in Fig.~1 that the ZB oscillations for~$n=0$ and~$\nu=\pm 1$ present mirror images
of each others. This means that, for some effects, the sum of both components can exactly cancel out.
This property is related to the fact that one can quantize the~$\hbar k_x$ momentum around the CNT circumference
in two equivalent ways: clockwise or counter-clockwise. In order to break this symmetry and get a nonzero
final result one can use an external magnetic field parallel to the tube's axis. The magnetic field gives
a preferential direction of the cyclotron motion that would break the above symmetry, see Ref.~\cite{Ando1993}.

To summarize, the density matrix formalism is used for the description of electron Zitterbewegung in carbon
nanotubes is order to include the dephasing processes. The decoherence of electrons in conduction and valence subbands,
as characterized by the time~$T_2$, quickens the decay of ZB oscillations, while the interband relaxation,
as described by the time~$T_1$, influences the rectilinear motion component. The decay of ZB due to
electron localization in space is of a different nature than that caused by the decoherence processes.


\begin{thebibliography}{99}
\bibitem{Schroedinger1930} E. Schrodinger, Sitzungsber. Preuss. Akad.
                          Wiss. Phys. Math. Kl. {\bf 24}, 418 (1930).
                          Schrodinger's derivation is reproduced in
                          A. O. Barut and A. J. Bracken, Phys. Rev. D {\bf 23}, 2454 (1981).
\bibitem{BjorkenBook}     J. D. Bjorken and S. D. Drell, {\it Relativistic Quantum Mechanics} (McGraw-Hill, New York,1964).
\bibitem{Lock1979}        J. A. Lock, Am. J. Phys. {\bf 47}, 797 (1979).
\bibitem{Gerritsma2010}   R. Gerritsma, G. Kirchmair, F. Zahringer, E. Solano, R. Blatt, and C. F. Roos,
                          Nature {\bf 463}, 68 (2010).
\bibitem{Zawadzki2005}    W.Zawadzki, Phys. Rev. B {\bf 72}, 085217 (2005).
\bibitem{Schliemann2005}  J. Schliemann, D. Loss, and R. M. Westervelt, Phys. Rev. Lett. {\bf 94}, 206801 (2005).
\bibitem{Zawadzki2011}    W. Zawadzki and T. M. Rusin, J. Phys.: Condens. Matter {\bf 23}, 143201 (2011).
\bibitem{Zawadzki2006}    W. Zawadzki, Phys. Rev. B {\bf 74}, 205439 (2006).
\bibitem{Rusin2007b}      T. M. Rusin and W. Zawadzki, Phys. Rev. B {\bf 76}, 195439 (2007).
\bibitem{Rusin2013b}      T. M. Rusin and W. Zawadzki, arXiv:1312.4709 (2013).
\bibitem{Ando1993}        H. Ajiki and T. Ando, J. Phys. Soc. Jpn {\bf 62}, 1255 (1993); {\bf 62}, 1270 (1993).
\bibitem{BoydBook}        R. W. Boyd, {\it Nonlinear Optics} (Academic Press, San Diego, 2003).
\bibitem{Kono2003}        J. Kono, G.N. Ostojic, S. Zaric, M.S. Strano, V.C. Moore, J. Shaver, R.H. Hauge, and R.E. Smalley,
                          Appl. Phys. A {\bf 78}, 1093 (2004).
\bibitem{Note2}           The ``naive'' calculation of the trace~$u{\rm Tr}(\hat{\rho}\sigma_x)$ is erroneous
                          since the two operators under the trace are in different representations.
\bibitem{Voisin2003}      J. S. Lauret, C. Voisin, G. Cassabois, C. Delalande, Ph. Roussignol, O. Jost, and L. Capes,
                          Phys. Rev. Lett. {\bf 90}, 057404 (2003).

\end{thebibliography}
\end{document}